# Regional Impacts of COVID-19 on Carbon Dioxide Detected Worldwide from Space

Brad Weir*[1,2], David Crisp[3], Christopher W O'Dell[4], Sourish Basu[2,5], Abhishek Chatterjee[1,2], Jana Kolassa[2,6], Tomohiro Oda[1,2], Steven Pawson[2], Benjamin Poulter[7], Zhen Zhang[8], Philippe Ciais[9], Steven J Davis[10], Zhu Liu[11], and Lesley E Ott[2]

[1]Universities Space Research Association, Columbia, Maryland, USA
[2]Global Modeling and Assimilation Office, NASA Goddard Space Flight Center, Greenbelt, Maryland, USA
[3]Jet Propulsion Laboratory, Pasadena, California, USA
[4]Cooperative Institute for Research in the Atmosphere, Colorado State University, Fort Collins, Colorado, USA
[5]Earth System Science Interdisciplinary Center, University of Maryland, College Park, Maryland, USA
[6]Science and Systems and Applications Incorporated, Lanham, Maryland, USA
[7]Biospheric Sciences Laboratory, NASA Goddard Space Flight Center, Greenbelt, Maryland, USA
[8]Department of Geographical Sciences, University of Maryland, College Park, Maryland, USA
[9]Laboratoire des Sciences du Climat et de l'Environnement, Gif sur Yvette, France
[10]Department of Earth System Science, University of California, Irvine, USA
[11]Department of Earth System Science, Tsinghua University, Beijing, China
* Correspondence to: brad.weir@nasa.gov

**Abstract**
Activity reductions in early 2020 due to the Coronavirus Disease 2019 pandemic led to unprecedented decreases in carbon dioxide ($CO_2$) emissions. Despite their record size, the resulting atmospheric signals are smaller than and obscured by climate variability in atmospheric transport and biospheric fluxes, notably that related to the 2019–2020 Indian Ocean Dipole. Monitoring $CO_2$ anomalies and distinguishing human and climatic causes thus remains a new frontier in Earth system science. We show, for the first time, that the impact of short-term, regional changes in fossil fuel emissions on $CO_2$ concentrations was observable from space. Starting in February and continuing through May, column $CO_2$ over many of the World's largest emitting regions was 0.14–0.62 parts per million less than expected in a pandemic-free scenario, consistent with reductions of 3–13% in annual, global emissions. Current spaceborne technologies are therefore approaching levels of accuracy and precision needed to support climate mitigation strategies with future missions expected to meet those needs.

**One Sentence Summary:** This paper shows, for the first time, that the regional impact of COVID-19 on atmospheric $CO_2$ was observable and quantifiable from space.

# Introduction

Reductions in human activity at the beginning of 2020 in response to the Coronavirus disease 2019 (COVID-19) produced the largest short-term change in fossil fuel and cement carbon dioxide ($CO_2$) emissions since the Industrial Revolution (*1*). Preliminary emissions estimates for 2020 based on economic activity data suggest that, compared to 2019 emissions, daily global emissions decreased by as much as 15–20% in April (*2*). Accumulated from the start of the year, these reductions reached ~7.8% by June (*3*) and are expected to total ~4% (low estimate) to ~10% (high estimate) for the year, with the exact annual decrease depending on the intensity of the reduction during the lockdowns, and the timing of the return of economic activity to pre-pandemic levels (*2*). Reductions in human activities were also indicated in satellite observed changes in nighttime light intensity (*4*), and short-lived, combustion-related pollutants, e.g., nitrogen dioxide ($NO_2$; *5–7*). While activity-based estimates are consistent with reductions in satellite $NO_2$ observations (*2*), the relationship of $NO_2$ to $CO_2$ emissions depends on combustion efficiency which varies significantly across sectors and regions. Furthermore, $CO_2$ emission estimates based on recent activity data, rather than the annual reported inventories typically used by "bottom-up" estimates, rely on different metrics and are thus subject to their own unique uncertainties. The two most well-known products (*2*, *3*), for example, intentionally produce estimates with significant day-to-day variability and would benefit from independent verification and analysis, e.g., by comparison to energy data (*8*), their spatiotemporal disaggregations (*9*), and the estimates that follows.

For the past two decades, space agencies from around the world have planned and launched several satellite missions to observe vertical column average $CO_2$ ($XCO_2$) with a long-term goal of quantifying anthropogenic $CO_2$ emissions and their trends. The current constellation includes Japan's Greenhouse Gases Observing Satellite (GOSAT; *10*), launched in 2009, NASA's Orbiting Carbon Observatory-2 (OCO-2; *11*, *12*) in 2014, Japan's GOSAT-2 (*13*) in 2018 and NASA's OCO-3, deployed in 2019 on the International Space Station (*14*). These missions were all designed as sounders that regularly sample the atmosphere at high precision, instead of mapping it in its entirety, with a strong focus on understanding the terrestrial biosphere. Future missions are expected to place an increasing focus on understanding anthropogenic emissions and improve coverage with greater swath widths and/or by sampling the atmosphere multiple times a day, e.g., NASA's Geostationary Carbon Observatory (GeoCarb; *15*) positioned over the Americas and many other ongoing international efforts (*16*).

Developing a system that uses atmospheric $CO_2$ observations to monitor changes in anthropogenic emissions remains a landmark achievement needed to support the implementation of international climate accords (*17*, *18*). Unlike $NO_2$ observations, which display clear plumes with high concentrations over emitting areas, $CO_2$ has a long lifetime in the atmosphere and is well-mixed. Furthermore, in any given month, regional terrestrial biospheric fluxes have similar or greater magnitudes than fossil fuel emissions. This means that the $CO_2$ signals caused by even large emissions changes are confounded by those from long range atmospheric transport and natural fluxes. To verify emissions changes with atmospheric $CO_2$ observations, the eventual goal is to sample the atmosphere as densely and frequently as possible above and downwind of emitting areas. This is not achievable with the current sparse surface network focused primarily on background $CO_2$ but becomes increasingly possible with satellite observations. Below, we present our approach for monitoring changes in atmospheric $CO_2$, analyze the observed changes in $XCO_2$ in 2020, and demonstrate that our system can detect and quantify the impact of COVID-19 on $XCO_2$, despite the significant difficulties noted in other studies (*19, 20, 21*). We conclude with a discussion of the scientific implications of those results.

## Results

**Monitoring $CO_2$ in Near Real Time**

The Goddard Earth Observing System (GEOS)/OCO-2 atmospheric carbon monitoring system has several unique characteristics that enable it to capture and quantify the signal due to COVID-19 (interactive visualizations available online at Refs. *22–24*). First, it takes advantage of the unique coverage and precision of OCO-2 measurements. Mixed throughout the atmosphere, a 7% reduction in annual fossil fuel emissions represents just a 0.33 ppm change (*25, 26*) against the global marine boundary layer background concentration of 412.22 ppm in January 2020 (*27*) assuming all other fluxes remain the same. While previous instruments have had insufficient coverage, accuracy, and/or precision to detect signals of this size, they remain within the nominal bounds of OCO-2 (*28, 29*). Second, it uses coupled meteorology-carbon cycle components within GEOS (*30*) and data assimilation to infer three-dimensional, gridded fields of $CO_2$ for the entire OCO-2 data record which can be averaged vertically and temporally as needed (see Materials and Methods and Figs. 1, S1–S3). By using a transport model, our approach accounts for the year-to-year variability in $CO_2$ due to differences in atmospheric circulation: even with no change in surface fluxes, transport variability can cause several ppm differences in $XCO_2$ over the same area from one year to the next (*31*) and one day to the next (Fig. 1B, C). This difference is especially relevant over North America, where passing weather systems cause sharp gradients across frontal boundaries (*32*). Analyses of $XCO_2$ retrievals that do not account for transport variability (*20, 21, 33*) are therefore unlikely to capture year-to-year differences in emissions, especially given the sparse and infrequent sampling of OCO-2 over emitting areas. Our approach calculates anomalies against a simulated baseline surface flux scenario with the given year's transport to account for known transport variability. Without this step, transport variability overwhelms the anomaly uncertainty (see Materials and Methods and Fig. S4). Finally, our system produces regular updates in near real time (NRT), taken here to mean a latency of less than a month, enabling the study of changes in the carbon cycle as they occur (*34*). Other common methods for inferring surface fluxes from atmospheric observations, e.g., flux inversion systems (*31, 35, 36*), typically trail the current date by several months or longer or are limited to a fixed period.

**Unprecedented $CO_2$ Anomalies in Early 2020**

Over much of the Northern Hemisphere, home to most of the World's largest economies and more than 95% of global total emissions, 16-day running means of $XCO_2$ from the GEOS/OCO-2 analysis show consistent, negative anomalies compared to a pandemic-free scenario (see Materials and Methods) beginning in February 2020 and continuing through May (Fig. 2). At the country/regional level, $XCO_2$ anomalies show a steep initial decline coinciding with the implementation of activity restrictions and a subsequent levelling off with the relaxation of those measures (Fig. 3). This phasing corroborates the finding from activity-data indicators that emissions dropped precipitously during the initial confinement and then slowly recovered or plateaued (*1–3*): a simulation of the expected 2020 fossil fuel anomaly using the daily, activity-based estimates of Ref. 3 is depicted with blue asterisks in Fig. 3. Overall, our results and the bottom-up simulation agree about the magnitude of reductions in $XCO_2$ growth at a country/regional level, with the analysis having slightly more temporal variability since it represents the anomaly from all fluxes, not just the fossil fuel component.

February–May 2020 anomalies over China, Europe, and the United States each exceeded the typical variability over the baseline period of 2017–2019 (see Materials and Methods): peak 1σ uncertainties ranged from 0.14–0.32 ppm, while peak reductions in $XCO_2$ growth reached 0.32–0.42 ppm (Table 1). By averaging those reductions and taking the geometric mean of their uncertainties, we find a 1σ range of 0.14–0.62 ppm for the Northern Hemisphere. Assuming the average reduction over the entire atmosphere is the same number at the end of the year, these estimates would produce 0.30–1.3 Pg C less of $CO_2$ in the atmosphere (*25*, *26*), corresponding to a 3–13% reduction in the 10 Pg C global fossil fuel emissions total estimated for 2019 (*37*). The conversion of ppm $CO_2$ to Pg C used above (*25*) is only a rough indicator of emissions, especially since interhemispheric mixing takes over a year to transport a signal from the Northern to Southern Hemisphere (*38*).

The monitored changes in $XCO_2$ over the Northern Hemisphere in February–May 2020 are primarily attributable to reductions in fossil fuel emissions for two reasons. First, late 2019 through 2020 saw neutral to weak La Niña conditions (*39*, *40*) of the El Niño/Southern Oscillation (ENSO). Globally, the annual growth rate of $CO_2$ correlates well with a linear combination of total anthropogenic emissions and the Niño 3 or 3.4 ENSO index (*41*, *42*). The latter term, which serves as a proxy for biospheric variability, is small in 2020 (*39*, *40*), indicating a strong anthropogenic role in the growth rate anomaly. Regionally and monthly, ENSO remains a dominant driver of biospheric anomalies, but not without notable exceptions (*43*). Second, the months of February–May occur during a "shoulder" season in which net biospheric exchange (NBE) is near its smallest (Figs. S5–S7), making it an ideal time to capture an anomaly driven by fossil fuel emissions. Transport simulations of 2020 anomalies from the Lund, Potsdam, Jena–Wald, Schnee und Landscaft (LPJ-wsl; *44*, *45*) and Catchment–Carbon and Nitrogen (Catchment-CN; *46*) terrestrial biosphere models (see Materials and Methods) also indicate that the biospheric anomalies in the Northern Hemisphere were relatively weak in February–May (Fig. S7).

One notable disagreement between the GEOS/OCO-2 analysis and the bottom-up simulation is in the timing of the reduction over the United States. In the bottom-up simulation, reductions in $XCO_2$ growth begin before activity restrictions as air with less $CO_2$ is transported from China, across the North Pacific, and eventually to the United States, a process that takes several days. These reductions are not apparent in the monitoring system. Over China (Fig. 3A) and the North Pacific (Fig. 2), where we expect to see sustained reductions in $XCO_2$, there is almost a complete rebound following the Lunar New Year. This is consistent with rebounds in $NO_2$ observations from satellites (*5*) and in situ sensors (*7*). While another study (*6*) found a rebound in $NO_2$ emissions following the Lunar New Year based on satellite observations, they did not find a complete recovery to pre-pandemic levels. There are several factors that could play a role in these discrepancies, each of which requires further investigation. In particular, uncertainties in Chinese emissions are greater than perhaps any other region (*47*, *48*), preventing us from making any strong conclusions about the magnitude of the recovery in their emissions. Nevertheless, these differences cannot be attributed to observational coverage or the data assimilation system alone—an observing system simulation experiment (OSSE) that samples the simulated values at the time and place of OCO-2 soundings and assimilates the result is able to reproduce simulated signals (Figs. S8, S9)—nor can they be linked to any significant anomalies in aerosol optical depth (Fig. S10), which is a common cause of retrieval error. Finally, companion simulations of biospheric anomalies suggest a small positive adjustment, over the North Pacific and United States (Fig. S7), although the difference is smaller than the within-model spread (indicated with stippling) and not great enough to account for the entire difference between the analysis and bottom-up simulation (yellow triangles, Fig. S8C). These results reinforce those of previous studies (*19*, *20*), which found it difficult to detect a COVID-19 signal over China using OCO-2 data.

While decreases in 2020 $XCO_2$ growth due to COVID-19 were apparent in the Northern Hemisphere, the same cannot be said of the Tropics and Southern Hemisphere, where biospheric variability complicated the interpretation of any COVID-19 signal. Starting in 2019 and continuing through February 2020, GEOS/OCO-2 captured another striking change in $XCO_2$, this time originating from the influence of a record-breaking climate anomaly on the terrestrial biosphere (Fig. 4). In 2020, well before their COVID-19 related restrictions, $XCO_2$ growth dropped over India and sub-Saharan Africa and increased over Australia (Fig. 5). During this period, countries surrounding the Indian Ocean were experiencing the tail end of the 2019–2020 Indian Ocean Dipole (IOD) whose Dipole Mode Index was in the greatest positive phase in recorded history (*39*, *49*), setting monthly (October 2019) and 3-month average (September–November 2019) all-time highs. The impact of the IOD on the terrestrial biosphere and atmospheric circulation began in 2019, when both sub-Saharan Africa and India had wetter than usual boreal autumns—during the positive phase, cooler-than-normal sea surface conditions persist in the eastern Indian Ocean with warmer-than-normal conditions in the western tropical Indian Ocean (*50–52*). This East-West contrast in ocean conditions alters the wind, temperature, and rainfall patterns in the region, typically bringing mild temperatures and floods to sub-Saharan Africa and the Indian subcontinent (*53*) and high temperatures and droughts to East Asia and Australia (*54*), among other ecological and socio-economic impacts. That increased rainfall over sub-Saharan Africa and the Indian subcontinent resulted in an extremely productive agricultural year and bumper crop harvests (*55*), while high temperature and drought conditions resulted in a record-setting fire season throughout Australia (*56*). The impact of these extremes on the carbon cycle persisted well into 2020, eventually falling off in early March (Figs. 4, 5; and the companion biospheric simulations in Figs. S7, S9).

## Discussion

We found that satellite monitored changes in early 2020 $XCO_2$ due to the COVID-19 pandemic were small (0.24–0.48 ppm), negative, and consistent with country-level activity data. The United States, Europe, and Asia each saw noticeable reductions in $XCO_2$ growth coinciding with restrictions on activity and a return to typical growth as those restrictions were eased. Attribution of these signals to changes in anthropogenic emissions remains challenging: interannual variability in transport and biospheric carbon-climate teleconnections both drive concentration changes many times greater than the record-setting changes in regional anthropogenic emissions due to COVID-19. For example, increased net vegetation growth in India and Africa and fires and respiration in Australia driven by the record-setting 2019–2020 IOD produced the greatest $XCO_2$ anomalies of early 2020. The ability to detect fossil fuel $CO_2$ emissions changes in the midst of such climate variability is a significant milestone toward the long-term goal of monitoring future emissions, especially given the planned increase in space-based observing capability. Nevertheless, land and ocean flux variations related to ENSO and IOD, and their related uncertainties, continue to limit our ability to monitor and understand changes in anthropogenic emissions. Attribution of $CO_2$ anomalies to individual surface flux components, and not their total, remains an active area of research with growing importance due to the societal need to reduce and monitor emissions. This effort will benefit in the future from improvements in terrestrial biospheric models, planned increases in space-based $CO_2$ observations with a greater emphasis on fossil fuel emissions from NASA's GeoCarb, Japan's GOSAT constellation, and Europe's $CO_2$ Monitoring mission, colocation with other remote-sensing observations (e.g., $NO_2$), and continued in situ measurement and scientific analysis of carbon isotopes, e.g., $^{14}C$ in $CO_2$ data (*57*).

# Materials and Methods

**Data assimilation**

The GEOS/OCO-2 atmospheric carbon monitoring system tracks changes in global atmospheric $CO_2$ every three hours by ingesting OCO-2 Build 10 $XCO_2$ full-physics retrievals (*58*, *59*) into GEOS using a statistical technique commonly referred to as data assimilation (DA; used here) and/or state estimation (*60*, *61*). It has been previously documented for an earlier version of OCO-2 data (*12*) and available to the public on the NASA/ESA/JAXA trilateral Earth Observing Dashboard (*22*) and NASA COVID-19 Dashboard (*23*) since July 2019.

DA synthesizes simulations and observations, adjusting the state of atmospheric constituents like $CO_2$ to reflect observed values, thus gap-filling observations when and where they are unavailable. These features are particularly appealing given the narrow, 10 km wide swath and 16-day repeat time of OCO-2 (Figs. 1, S2). DA can be considered a machine learning (ML) method. Compared to interpolation, Kriging, and most other ML approaches, DA has the advantage that it makes estimates based on our collective scientific understanding of Earth's carbon cycle, as encapsulated within GEOS, rather than relying on functional relationships that rarely hold in nature. The value of relying on forecasted fields instead of functional relationships in data analysis has been understood in the numerical weather prediction community since at least 1954 (*62*), even before E. Lorenz's seminal work (*63*), yet receives less attention in other disciplines. Fig. 1 demonstrates the impact of data assimilation on OCO-2 coverage for April 2020. Before assimilation (top panel), there are notable patches of missing data where either clouds (e.g., the Amazon), aerosols (China), and high solar zenith angles (the poles) prevent reliable measurements. Assimilation produces three-dimensional $CO_2$ fields with global coverage that are updated every three hours (middle panels and Fig. S1). Values in missing areas are inferred from nearby observational data and model relationships. The 16-day running means (bottom panel) and monthly means analyzed in this paper follow from simple averaging of the assimilated $CO_2$ fields.

GEOS/OCO-2 uses the GEOS Constituent Data Assimilation System (CoDAS), a high-performance computing implementation of Gridpoint Statistical Interpolation (GSI; *64*), a technique for finding the analyzed state that minimizes the three-dimensional variational (3D-Var) cost function formulation of the differences between observed and simulated values. GEOS CoDAS ingests column retrievals of trace gas abundances, accounting for both their vertical sensitivity (i.e., averaging kernel) and a priori assumptions. While current versions of GSI include the ability to use four-dimensional variational (4D-Var) and hybrid ensemble-variational formulations (65), this application relies on the simpler 3D-Var technique. In GEOS CoDAS, the atmospheric circulation is constrained by the millions of remote sensing and in situ observations every hour included in the Modern Era Retrospective analysis for Research and Application, version 2 (MERRA-2; 66). This accurate representation of transport patterns at fine spatial resolutions is critical for interpreting measured variations that reflect a combination of nearby and distant surface fluxes due to the long lifetime of $CO_2$ and helps us reproduce atmospheric observations with high fidelity in the marine boundary layer (34) and over North America, where there is a wealth of data, e.g., airborne in situ measurements from NASA's Atmospheric Carbon and Transport–America (ACT-America) campaign (67). Extensive evaluation against these data, which are withheld from the assimilation, makes us confident in the ability of GEOS/OCO-2 to estimate regional signals with small magnitudes. In other applications, GEOS CoDAS has been used to analyze multi-decadal trends of stratospheric ozone (68) and the anomalously small ozone hole of 2019 (69).

All GEOS CoDAS runs in this paper use a similar methodology and setup to that described in Ref. 69 and the references therein. The horizontal grid has a nominal resolution of 50 km and there are 72 vertical levels starting from the surface, where they follow the terrain, and extending up to 0.01 hPa, where they follow fixed pressure values. The assimilation system processes observations in 6-hour intervals. At the beginning of each interval, it uses the GEOS model to simulate a 6-hour forecast/background and saves output every 3 hours for the purpose of time interpolation. It then solves for the minimum value of the cost function:

$$J(x) = \frac{1}{2}(x - x_b)^T B^{-1}(x - x_b) + \frac{1}{2}(y - Hx)^T R^{-1}(y - Hx),$$

where $x$ is the state vector of trace gas values at all grid points, $y$ is the observation vector, $H$ is the observation operator, $B$ is the background error covariance matrix, and $R$ is the observation error covariance matrix. This formulation abuses notation slightly as the GSI 3D-Var formulation assumes that the same increment $x - x_b$ is constant throughout the 6-hour interval, while using 3-hourly temporal interpolation for the evaluation of $Hx$. GEOS/OCO-2 uses a homogeneous, horizontally isotropic background error covariance $B$ whose diagonal is 0.15 ppm everywhere with a nominal horizontal error correlation length of 500 km and vertical error correlation length proportional to the vertical correlation length of the given tracer. The observation error covariance $R$ uses the reported retrieval error variances scaled by a factor of $0.85^2$ as its diagonal and has no cross-sounding correlations. While these crude error models could be improved, a posteriori diagnostics and evaluation against independent data indicate that the system is sufficiently well tuned. As an additional level of quality control, we do not assimilate retrievals that are over snow and ice, have a glint angle greater than 80°, or are in a swath with less than 4 footprints. Soundings with a reported uncertainty less than 0.001 ppm are also flagged and not assimilated. Cross-track variability of $XCO_2$, accounting for the retrieval mode and surface type, is included in the retrieval errors by geometrically averaging it with the reported values. The final step for each interval is to rerun the 6-hour forecast with the optimal increment $x^* - x_b$, where $x^*$ minimizes the cost function $J$, applied as a forcing in the same manner as for the meteorological variables (65).

Data processing is divided into six separate streams covering 2015–2020. Each stream begins on October 31 of the previous year to allow some equilibration of the analysis prior to the period of interest beginning on January 1. Differences between overlapping streams are less than 10% of the magnitude of the anomalies analyzed in this paper, and thus can be safely ignored. The results presented here use no $CO_2$ data other than OCO-2 observations in the present year, here 2020. In previous years, it uses a single number, the atmospheric growth rate, to set the global flux budget as described below.

A unique feature of GEOS/OCO-2 is its ability to process data in NRT, as retrievals become available to assimilate. This is accomplished primarily through the use of a surface flux collection, the Low-order Flux Inversion (LoFI; *34*) with distinct modes for retrospective and NRT simulations. In retrospective simulations, the system uses surface fluxes informed by several remote sensing datasets that include fire radiative power, nighttime lights, and vegetation properties like leaf area index (*30*) and atmospheric growth rate estimates derived from surface observations. In near real time, before many of these datasets become available, LoFI uses fluxes and a projected atmospheric growth rate based on data from previous years and the current ENSO phase (*41*, *42*). This dual capability ensures a strong, multi-platform data constraint on $XCO_2$ on previous years for computing anomalies while the products for the current year highlight areas where land, ocean, and fossil fuel fluxes deviate from expectations. For fossil fuel emissions, we use the 2018 version of the Open-source Data Inventory for Anthropogenic $CO_2$ (ODIAC; *70*), which estimates emissions by tracking fossil fuel consumption (i.e., barrels of oil, tons of coal, etc.) and cement production (*9*) and ends in 2018. For 2019, we rescale the 2018 monthly gridded maps to match the Global Carbon Project (GCP) 2019 global emissions estimate (*37*), and for 2020 we simply repeat 2019 emissions.

**Anomalies and pandemic-free baseline atmospheric $CO_2$ fields**

Even after constructing gap-filled $XCO_2$ maps, defining 2020 anomalies for $CO_2$ is more challenging than for most other species. For $NO_2$, which is short-lived, simply subtracting a multi-year climatological mean from the 2020 values is often sufficient for highlighting recent emissions changes (*5*), although recent research suggests meteorological variations can play a significant role in the interpretation of $NO_2$ changes (*7*, *71*). For $CO_2$ and other long-lived species, anomalies calculated against a climatological baseline reveal a strong imprint of circulation anomalies, which can have a greater impact than, and obscure the spatial signature of emissions changes.

To minimize the circulation influence, at the beginning of each year we start a companion GEOS simulation that is identical to the analyzed product, except that OCO-2 data are not assimilated. By subtracting the simulated anomaly from the analysis anomaly, we isolate the flux-driven signal observed by OCO-2 from the transport-variability-driven signal. We refer to this difference as the "analysis correction." The pandemic-free, baseline scenario is then the average of all analysis corrections for 2017–2019 plus the GEOS simulation for 2020. This represents 2020 transport while applying the mean analysis corrections due to assimilating OCO-2. Subtracting the pandemic-free field from the 2020 analysis then gives the flux driven GEOS/OCO-2 anomaly depicted in Figs. 2–5 and in the supplemental figures. Fig S4 depicts the difference between this anomaly calculation, which we call "transport aware", and an anomaly calculation that uses a simple climatology of previous years as a baseline. By not accounting for year-to-year transport variability, the latter has a much greater standard deviation across years, as seen in the increased stippling in Fig. S4.

We omit 2015 and 2016 from our baseline years because they contain one of the strongest ENSOs on record and are not representative of 2020, which was neutral in the first three months of 2020, and transitioned to a moderate La Nina in April 2020 (*38*, *39*). Strong ENSO signals produce significant inter-annual variability in carbon fluxes over ocean and land (*35*, *72*) as well as atmospheric circulation patterns (*73*). Figs. S11 and S12 add the 2015 and 2016 anomalies onto the plots from Figs. 3 and 5. The ENSO years (red) are clear outliers, supporting their exclusion from the analysis.

**Uncertainty quantification**

As an indicator of uncertainty, we use the range of analysis corrections for individual years in 2017–2019 depicted as the gray shading in Figs. 3 and 5. From the ranges, we calculate the $2\sigma$ uncertainty as half the min-to-max range of the gray shaded area, corresponding to an assumption that the 2017–2019 range represents about 95% of year-to-year variability in neutral ENSO conditions. The uncertainty ranges reported in Table 1 are consistent with evaluations of GEOS/OCO-2 against independent surface in situ and remote sensing observations and a posteriori tests of the statistical consistency of the data assimilation system (see Supplement). They are smaller than, but the same order as the errors reported by the analyses in several previous studies (*19*, *20*, *74–76*). This is to be expected as our uncertainty estimate does not include persistent biases, while those estimates do. They also coincide with a 0.15 ppm standard deviation of the analysis error uncertainty for the GEOS/OCO-2 fields calculated from an a posteriori diagnostic (*77*).

**Separating the COVID-19 atmospheric $CO_2$ signal from natural variability**

In order to help separate anthropogenic from natural variability, we perform two supplementary GEOS $CO_2$ simulations. The first transports the difference in 2020 and 2019 emissions from the daily activity-based fossil fuel estimates (*3*) through the atmosphere using the same settings as the GEOS/OCO-2 assimilation run (monthly global maps in Fig. S7). For these simulations, daily country-level emissions totals for 2019 and 2020 are spatially downscaled using 2015 monthly EDGAR v5.0 sector totals (*78*) for power generation, ground transportation, industry, aviation, residential energy usage, and international shipping. The second simulation aims to represent the difference in 2020 biospheric flux by transporting the difference between 2020 and the 2017–2019 average calculated using the LPJ-wsl dynamic global vegetation model (Figs. S5–S7). While LPJ-wsl is a different model of the terrestrial biosphere than we use for our prior fluxes, it is useful as a prognostic, independent method of identifying regional biospheric anomalies and has been demonstrated to realistically reproduce interannual variations in global net flux (*37*). For consistency, we apply the same MERRA-2 meteorological data used to force our transport simulations and as inputs to LPJ-wsl (*45*) and Catchment-CN (*46*).

**Acknowledgements:** This article is dedicated to the memory of Michael Freilich, the former director of the Earth Science Division of NASA. His skill, dedication, bravado, and commitment were critical to ensuring that the Orbiting Carbon Observatory-2 even existed. The authors would also like to thank Kenneth W. Jucks for his invaluable scientific insights and guidance as program manager, Peter H. Jacobs, Lori K. Perkins, Nikolay Balashov, and Frédéric Chevallier for their comments and suggestions. High-performance computing resources were provided by the NASA Center for Climate Simulation (NCCS). **Funding:** This work was supported by NASA's Carbon Monitoring System (NNH16DA001N-CMS 16-CMS16-0054), Science Team for the OCO Missions (NNH17ZDA001N-OCO2 17-OCO2-17-0010), and Modeling, Analysis, and Prediction (NNH16ZDA001N-MAP 16-MAP16-0165) projects. Some of the work described here was performed at the Jet Propulsion Laboratory, California Institute of Technology, under contract to the National Aeronautics and Space Administration. Government sponsorship acknowledged. **Author contributions**: All authors contributed to the development of the ideas described within this manuscript, the data collection, and the manuscript's composition. B. Weir led the writing and development of the data assimilation system. D. Crisp and C. O'Dell led the development and processing of OCO-2 retrievals. T. Oda led the development of the ODIAC fossil fuel emissions, B. Poulter and Z. Zhang the LPJ-wsl biospheric model, J. Kolassa the Catchment-CN model, and P. Ciais, Z. Liu, and S. Davis the NRT activity-based fossil fuel emissions. **Competing interests:** The authors declare no competing interests. **Data and materials availability:** All data needed to evaluate the conclusions in the paper are present in the paper. All data needed to reproduce the figures and tables in the paper are available at https://doi.org/10.5281/zenodo.5213009. OCO-2 data were produced by the OCO-2 project at JPL and were obtained from the free data archive maintained at the Goddard Earth Sciences Data and Information Services Center (GES DISC; https://disc.gsfc.nasa.gov/OCO-2).

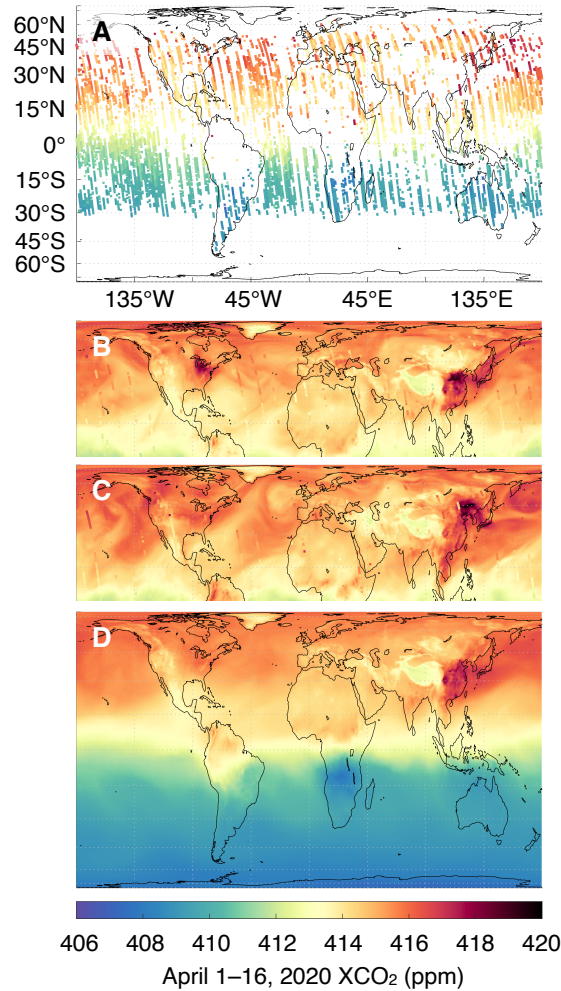

**Fig. 1. Snapshots of OCO-2 XCO$_2$ soundings and assimilated GEOS/OCO-2 fields.** (**A**) 16 days of OCO-2 XCO$_2$ soundings on April 1–16, 2020, (**B, C**) daily mean GEOS/OCO-2 XCO$_2$ at the beginning, April 1, 2020 (B) and end, April 16, 2020 (C) of the 16-day period with 8 days of assimilated OCO-2 data overlaid on each, and (**D**) the 16-day average of assimilated GEOS/OCO-2 XCO$_2$ over the same period. Data assimilation combines satellite observations (A) with a weather-resolving atmospheric model (B, C) to form gridded, time-varying, three-dimensional fields (Fig. S1), from which averages (D) and uncertainties (Figs. 2–5) follow. Since it accounts for the several ppm changes in the Northern Hemisphere from (B) to (C) due to meteorological and sub-monthly flux variability, the assimilation system can detect and quantify the much smaller COVID-19 signal (see Materials and Methods; Fig. S4). Monthly OCO-2 coverage and assimilated fits to data are depicted in Fig. S2, S3.

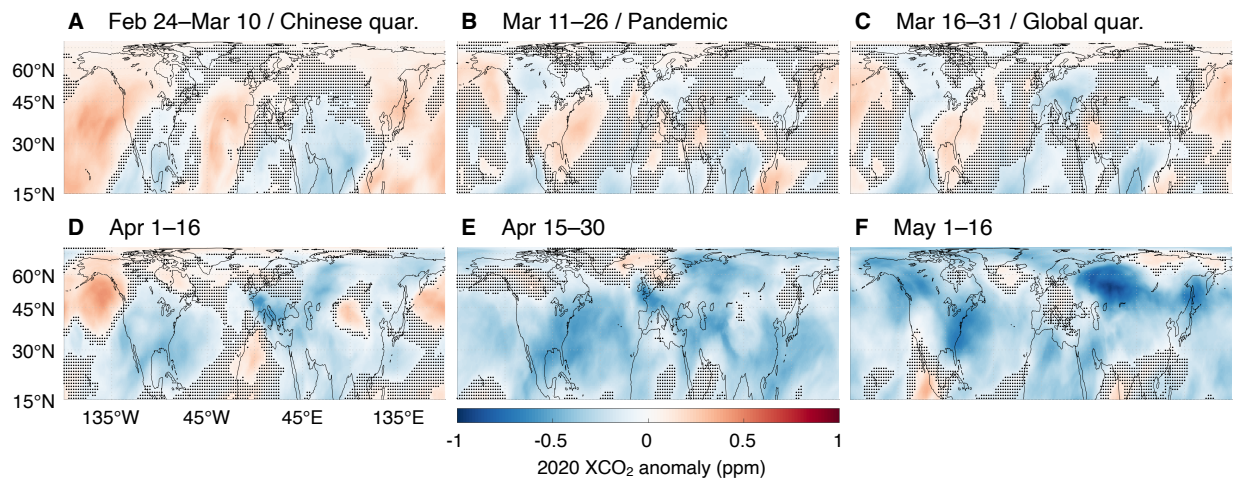

**Fig. 2. Spatial maps of GEOS/OCO-2 16-day moving average XCO$_2$ anomalies over the Northern Hemisphere from late February through May 2020.** Blue colors indicate decreases in XCO$_2$ growth compared to a pandemic-free scenario (see Materials and Methods), while red colors indicate increases. Stippling indicates points where the signal is less than half a standard deviation of the uncertainty. Reductions surpassing 1 ppm, depicted in deep blue, developed over North America and Europe in mid-March through May (**C**–**F**) as COVID-19 related restrictions on activity were put in place. Afterwards, in late May to early June, mixing by atmospheric transport, rebounds in emissions, and variability in the terrestrial biosphere diminish the magnitude and coherence of the COVID-19 signal (Figs. 2, S8).

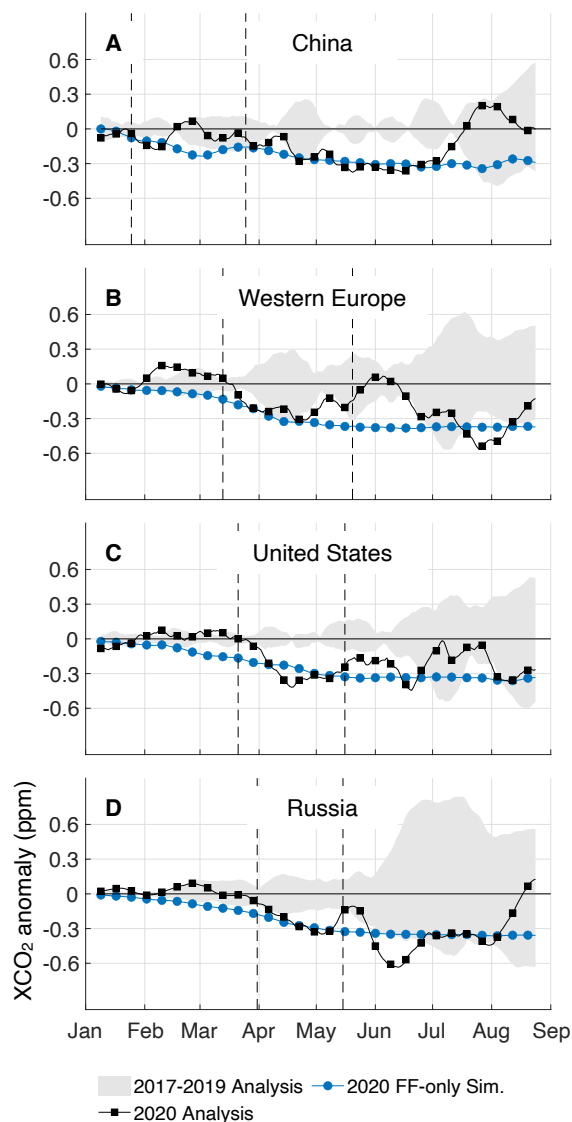

**Fig. 3. Time series of regional GEOS/OCO-2 16-day moving average XCO₂ anomalies.** (**A**) China, (**B**) Western Europe, (**C**) the United States, and (**D**) Russia. The solid black line with boxes indicates the 2020 anomaly, the gray shaded area the spread of anomalies across the baseline years (2017–2019), and the blue circles transport simulations for 2020 using activity-based fossil fuel emissions estimates (*3*). Dashed lines mark a rough beginning and end (when appropriate) to confinements for each area and are provided in Table 1. For additional simulations and analysis including histograms of daily sounding counts, see Fig. S8.

**Table 1. GEOS/OCO-2 February–May regional reductions in $XCO_2$ growth and associated uncertainties.** Reductions and uncertainties are calculated from the data depicted in Fig. 3 as the peak 2020 reduction (solid boxes) and peak $1\sigma$ 2017–2019 uncertainty (gray shading) during February–May (see Materials and Methods). Start dates and end dates are taken from activity data (*3*). The average reduction over all four regions is 0.38 ppm and average uncertainty is 0.24 ppm, giving a $1\sigma$ range of 0.14–0.62 ppm for the reduction over the Northern Hemisphere.

|  | Peak reduction | Peak $1\sigma$ uncertainty | Start | End |
|---|---|---|---|---|
| **China** | 0.37 ppm | 0.26 ppm | January 25 | March 25 |
| **Western Europe** | 0.32 ppm | 0.32 ppm | March 13 | May 20 |
| **United States** | 0.42 ppm | 0.14 ppm | March 21 | May 16 |
| **Russia** | 0.41 ppm | 0.22 ppm | March 31 | May 15 |

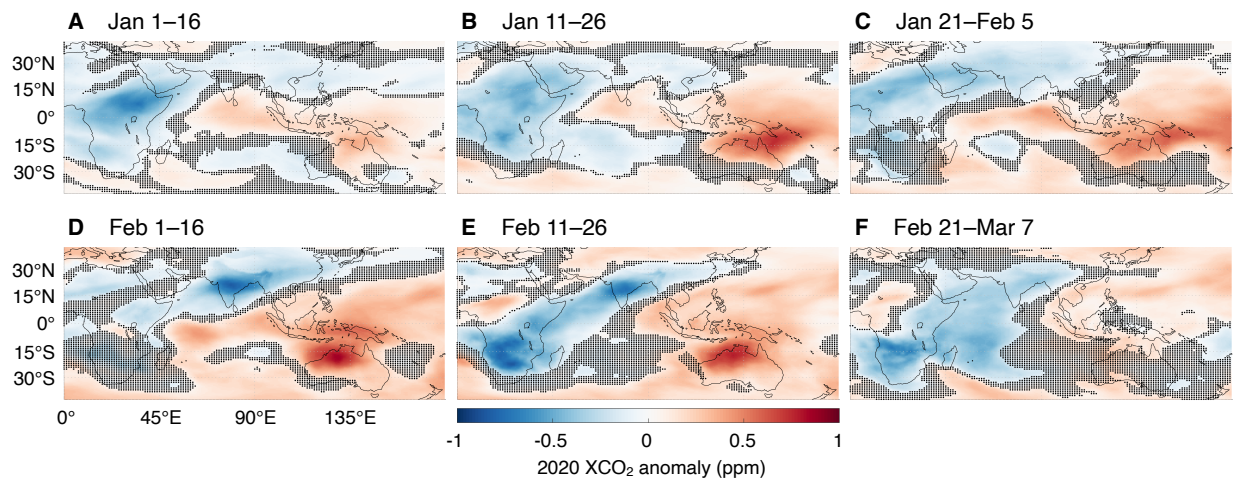

**Fig. 4. Identical to Fig. 2, but over the Indian ocean before the start of COVID-19 related confinements.** In contrast to Fig. 2, the dominant signal shown here is from the carbon-climate teleconnection between the 2019–2020 Indian Ocean Dipole (IOD), the strongest on record, and terrestrial biospheric exchange. In January (**A**–**C**) and February (**D**–**F**) 2020, there was increased biospheric uptake over India and Africa (blue colors) due to greater than average precipitation in the preceding months, while there was increased respiration and biomass burning over Australia and Southeast Asia (red colors) due to greater than average temperatures.

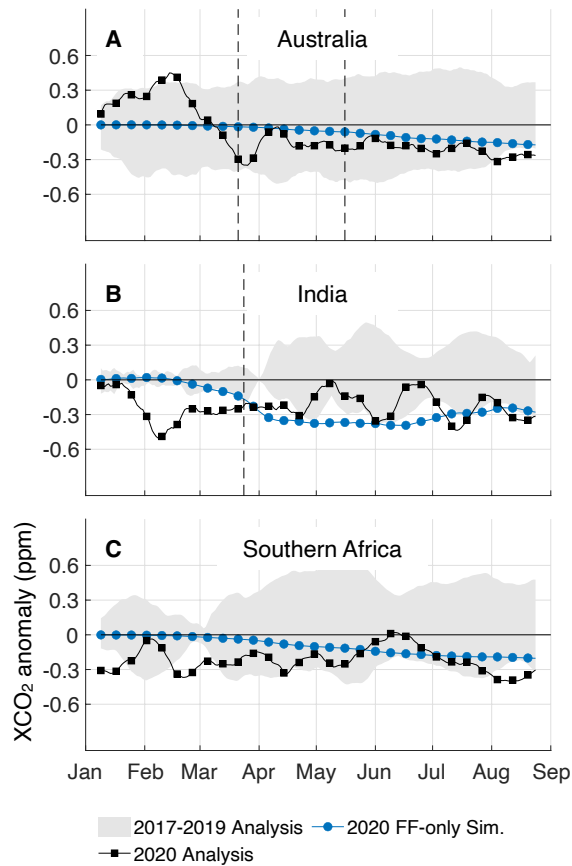

**Fig. 5. Identical to Fig. 3, but for different regions.** (**A**) Australia, (**B**) India, and (**C**) Southern Africa. The dominant signal is that of the IOD impact over India, but most of the anomalies are within the range of typical changes. As opposed to the Northern Hemisphere (Fig. 3), early in the calendar year is a time of significant biospheric activity in the Tropics and Southern Hemisphere (Figs. S5–S7), complicating the interpretation of any anthropogenic variability. For corresponding histograms of daily sounding counts, see Fig. S9.

# Supplementary Materials

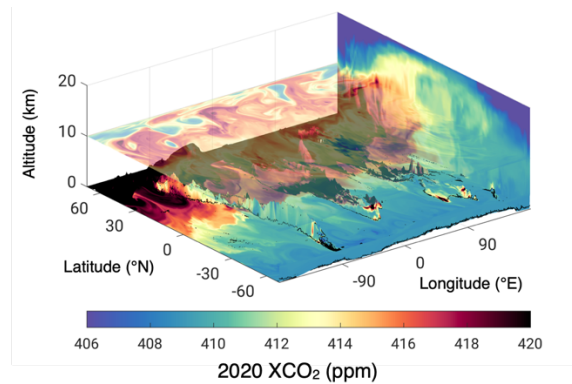

**Fig. S1. Transects of instantaneous GEOS/OCO-2 CO$_2$ on April 9, 2020 at 00:00 UTC at the surface (bottom), 10 km above sea level (top, transparent), and along the International Dateline (right)**. By reproducing the global, high-resolution, vertical and temporal variability of CO$_2$, the assimilation system can synthesize heterogenous data types across drastically different scales, e.g., satellite retrievals and in situ measurements from surface stations and aircraft.

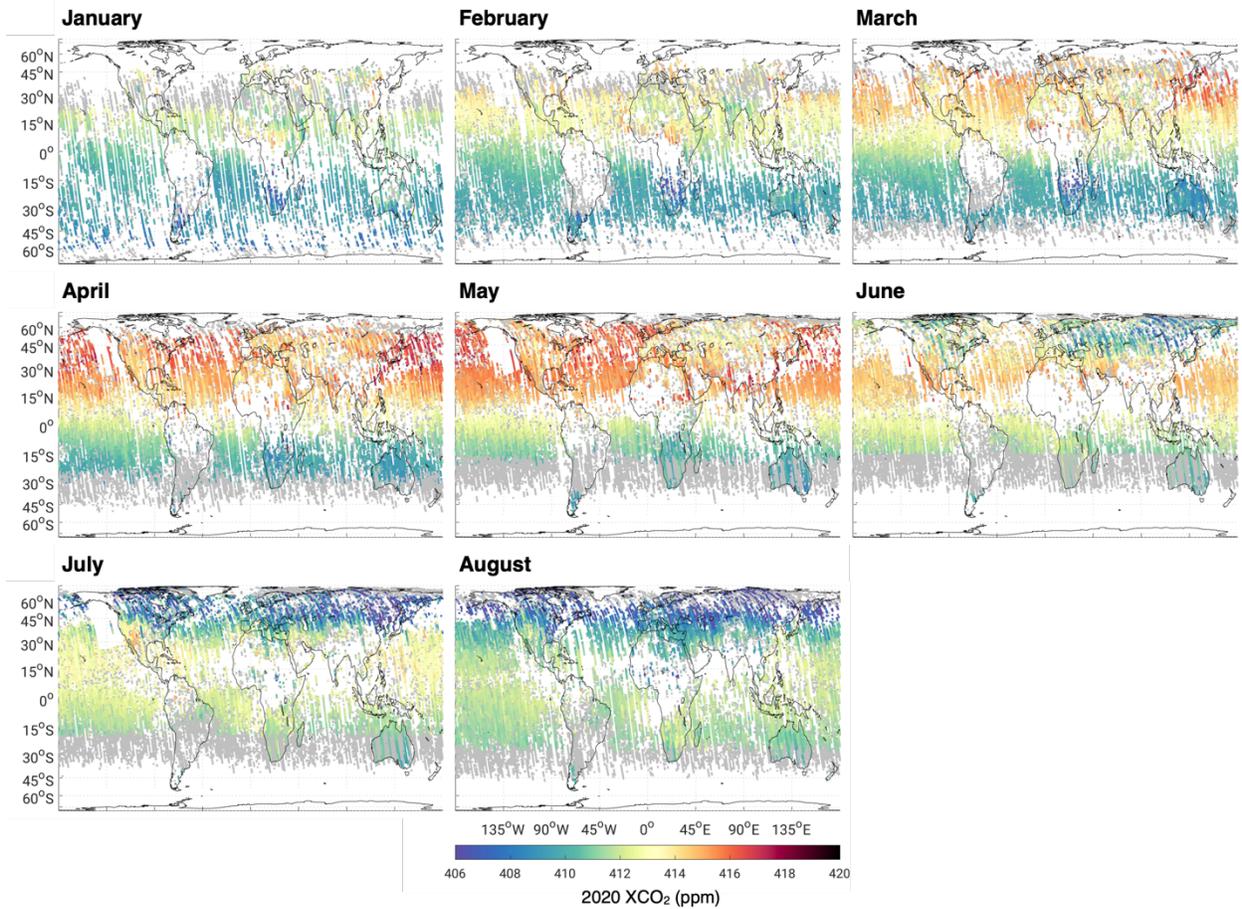

**Fig. S2. Monthly maps of OCO-2 B10 XCO$_2$ retrieval coverage.** The OCO-2 satellite flies in a sun-synchronous, low-Earth orbit with a local overpass time of 1:30 PM in NASA's Afternoon Train (A-Train) formation. Soundings flagged by our additional level of quality control are depicted in gray (see Materials and Methods). Its most significant effect is on coverage at high solar zenith angles over the ocean and Southern Hemisphere land.

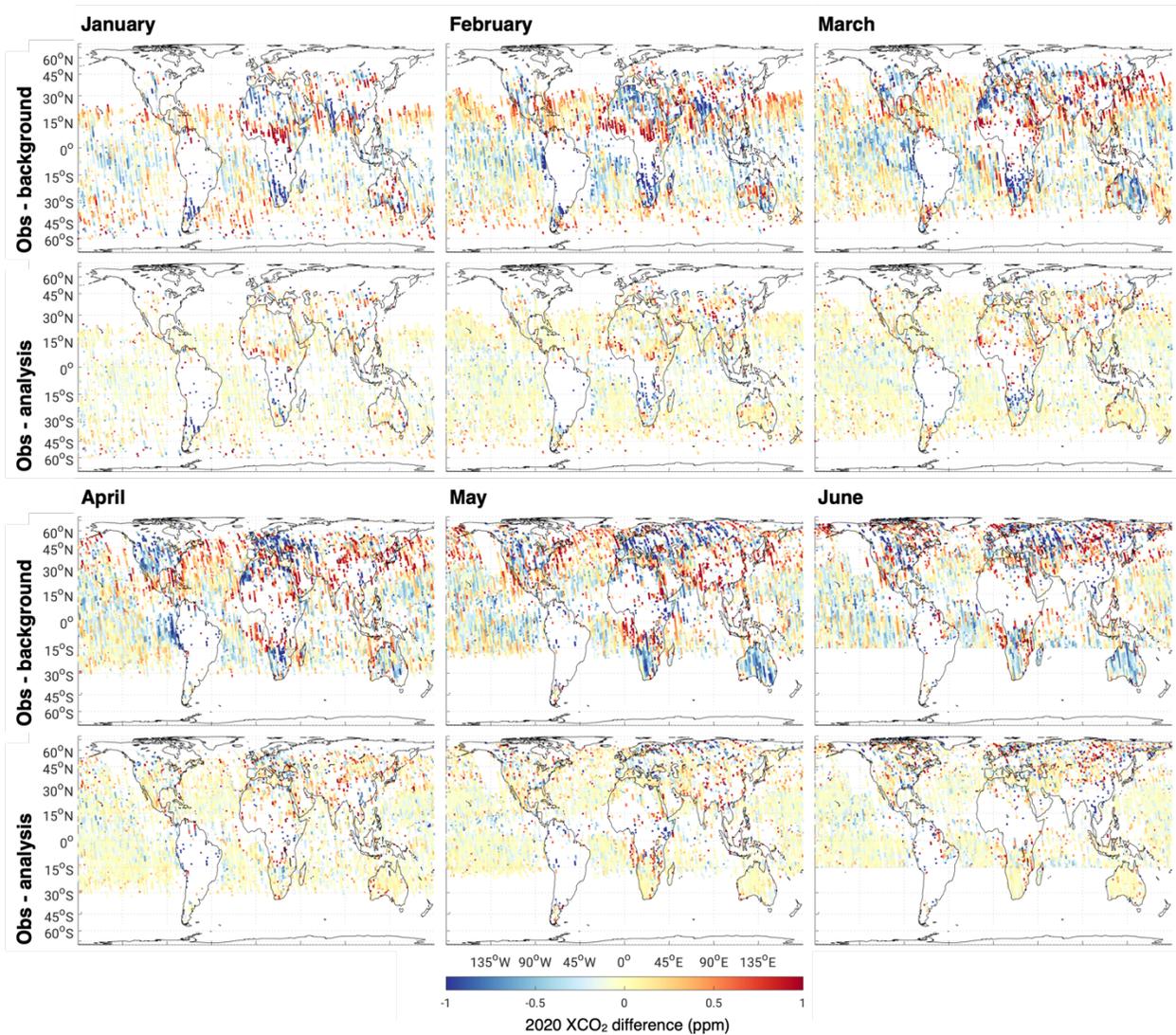

**Fig. S3 (continued below ...). Monthly maps of OCO-2 minus GEOS/OCO-2 background (before assimilation; top rows) and analysis (after assimilation; bottom rows) $XCO_2$.** Assimilation clearly improves the fits to the assimilated data, as intended. Differences after assimilation have O(0.1 ppm) magnitudes, further supporting the uncertainty quantification in the paper. Only assimilated soundings, which pass the additional quality filters (see Materials and Methods) and are thus depicted in color in Fig. S2, are shown here. A final layer of quality control, which discards data whose absolute difference with the model exceeds 3.5 standard deviations of the observation error, accounts for sporadic data points depicted in Fig. S2 and missing here.

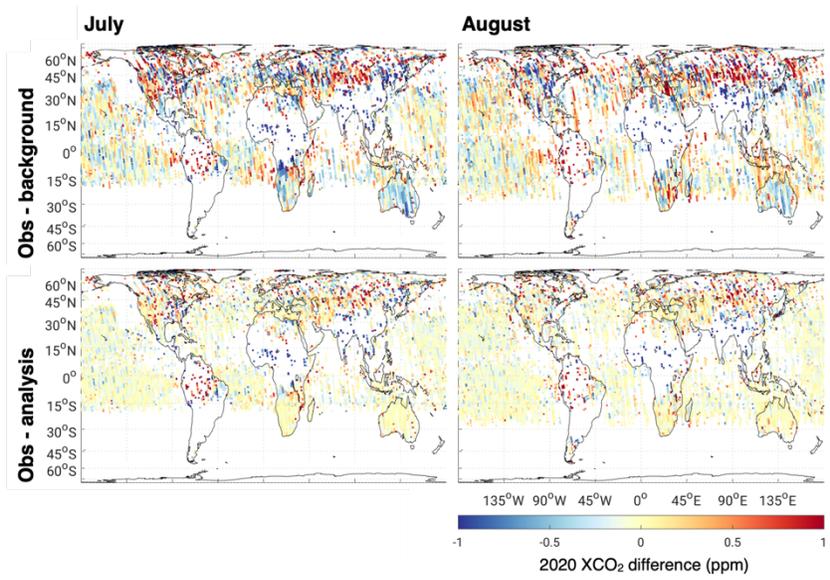

**Fig. S3 (... continued from above).**

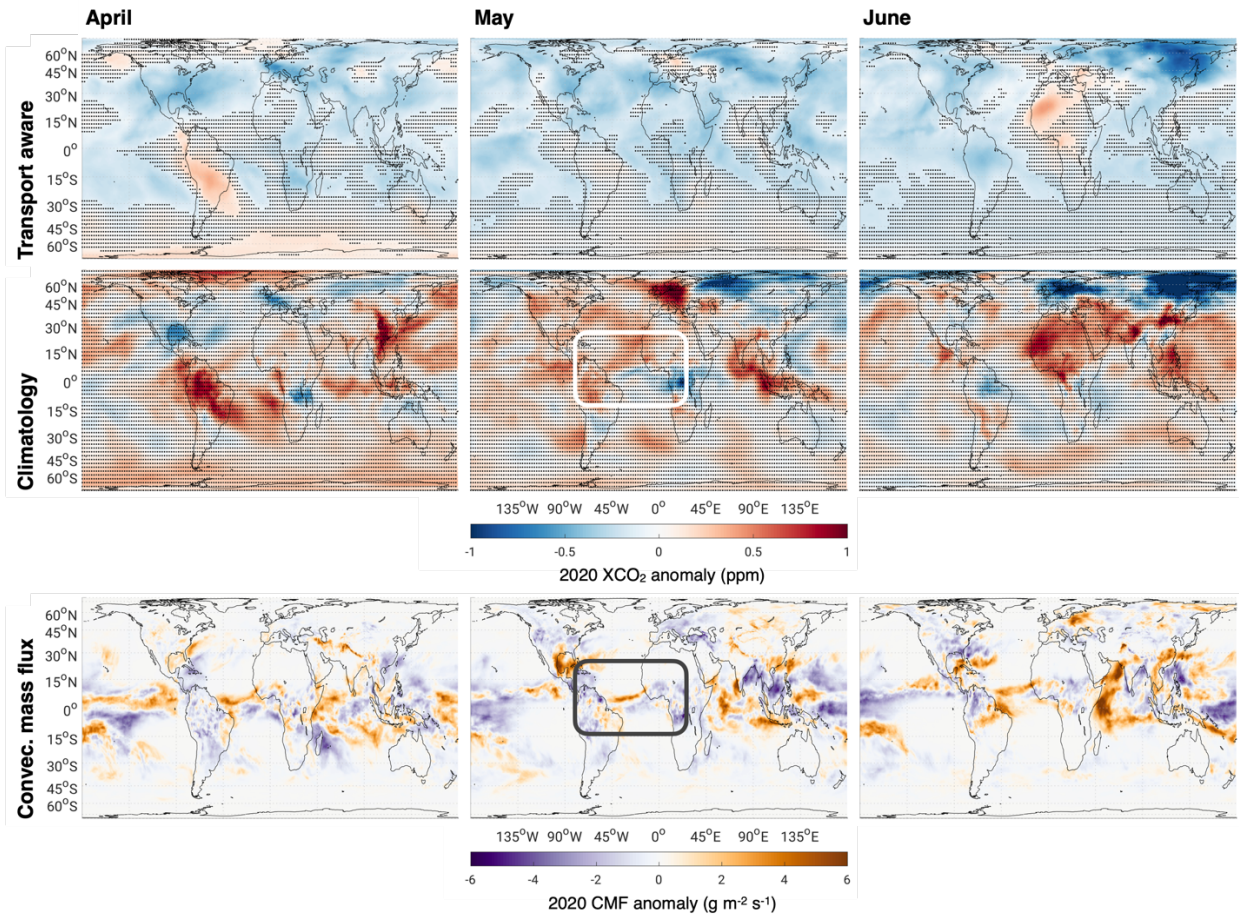

**Fig. S4. Anomalies of GEOS/OCO-2 monthly average XCO$_2$ during early 2020 computed with two different pandemic-free baseline scenarios along with anomalies in convective mass flux from MERRA-2.** The first row, used in the text, computes a baseline that accounts for year-to-year variability in atmospheric transport by subtracting out simulated values. The second row computes a baseline that is the climatology of previous years with a constant offset and makes no adjustment for inter-annual variability in transport. Stippling in both indicates when the signal is less than half a standard deviation. By not accounting for transport, the latter anomaly estimate produces a signal with far greater uncertainty, so much so that every point is stippled. In addition, the large positive/negative anomaly over the tropical Atlantic in May 2020 (inset box) is a direct result of year-to-year changes in convective mass flux (bottom row) and not due to a surface flux anomaly, which we hope to detect. Similar plots for anomalies in sea-level pressure and convective mass flux for each month are given in Fig. S12.

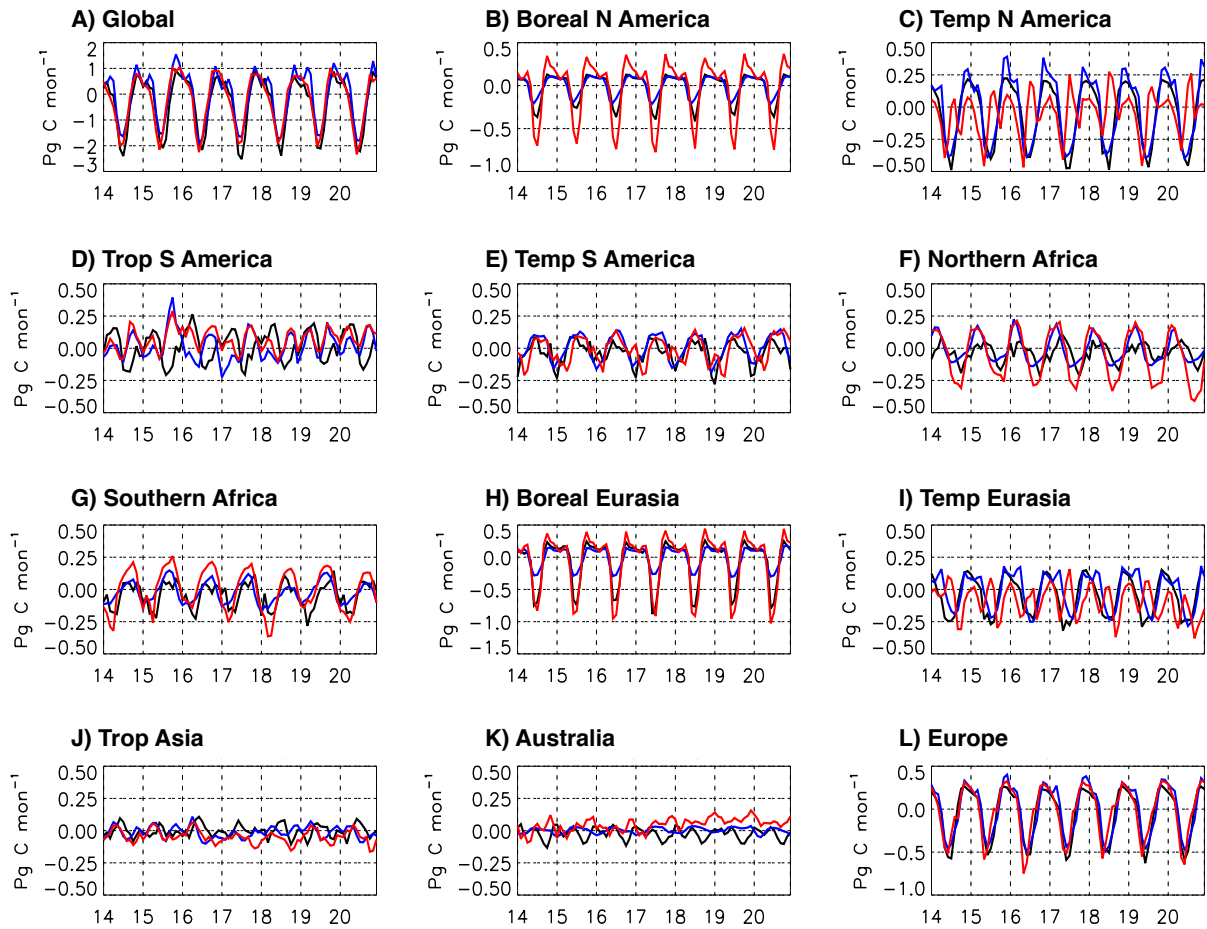

**Fig. S5. NBE from our prior surface fluxes (LoFI, black), the LPJ-wsl model (red), and the Catchment-CN model (blue) used for the biospheric anomaly simulation.** While the models reproduce roughly consistent global seasonal cycles (A), LPJ-wsl has a seasonal cycle that is about twice as great as LoFI in Boreal North America, North Africa, and South Africa, and Catchment-CN has a seasonal cycle about twice as small as LoFI and LPJ-wsl in Boreal Eurasia. These differences suggest a COVID-19 signal in $XCO_2$ is most likely to be detected in February–May before the greatest uncertainties among the models occur.

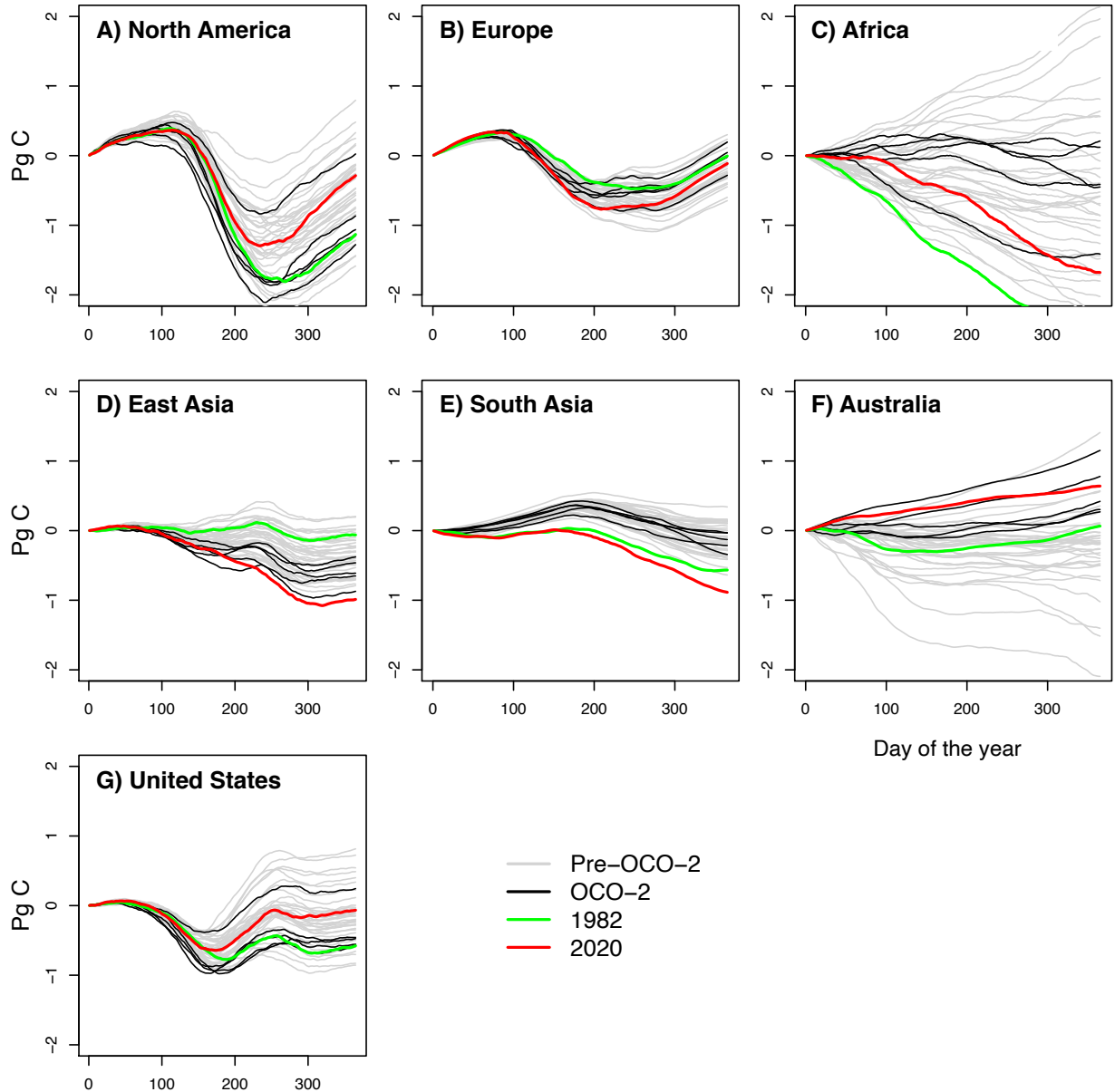

**Fig. S6. Cumulative NBE anomalies from LPJ-wsl terrestrial biospheric model over different countries/regions.** Each line represents a different year with years before the OCO-2 launch (2014) drawn in gray, years after the OCO-2 launch drawn in black, 1982 drawn in green (most recent IOD of comparable magnitude), and 2020 drawn in red. The year 2020 is a clear outlier in South Asia and Oceania/Australia, indicating strong biospheric anomalies. Over North America, Europe, and East Asia (viz., China), 2020 is a typical year for the biosphere, if not more of a source (positive values) of carbon to the atmosphere. The rapid increase in inter-annual variability in the Northern Hemisphere around the end of May (day 150), shows that a fossil fuel anomaly is best observed during its winter and spring.

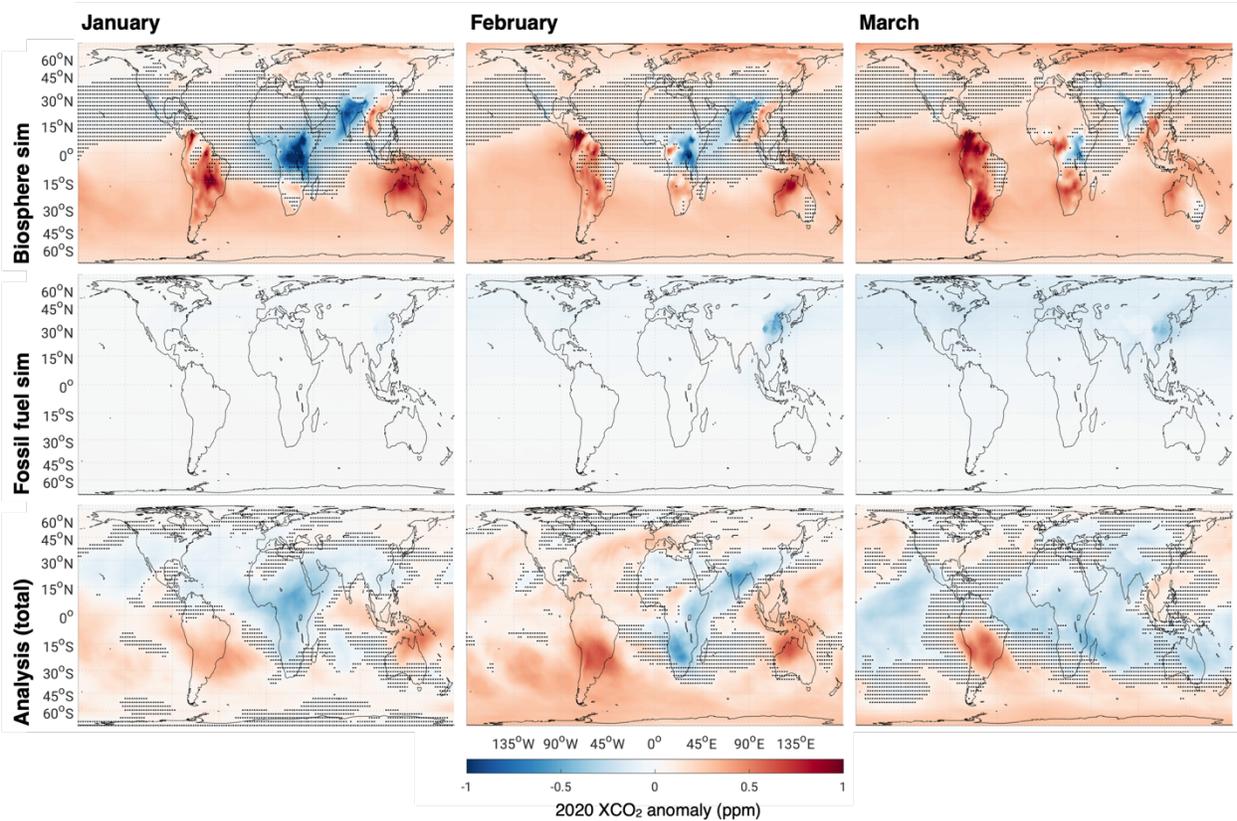

**Fig. S7 (continued below ...). Monthly maps of XCO$_2$ in ppm from biospheric anomaly simulations (top), fossil fuel anomaly simulation (middle), and GEOS/OCO-2 analysis anomaly (bottom) with stippling indicating signals smaller than half a standard deviation.** In January through February, the GEOS/OCO-2 anomaly captures the IOD signal over India, Australia, and Africa, showing remarkable agreement with the spatial patterns from the biospheric simulations. The amplitudes of those patterns are nevertheless quite different, with LPJ-wsl estimating a seasonal amplitude more than twice that of LoFI in some regions (Fig. S5).

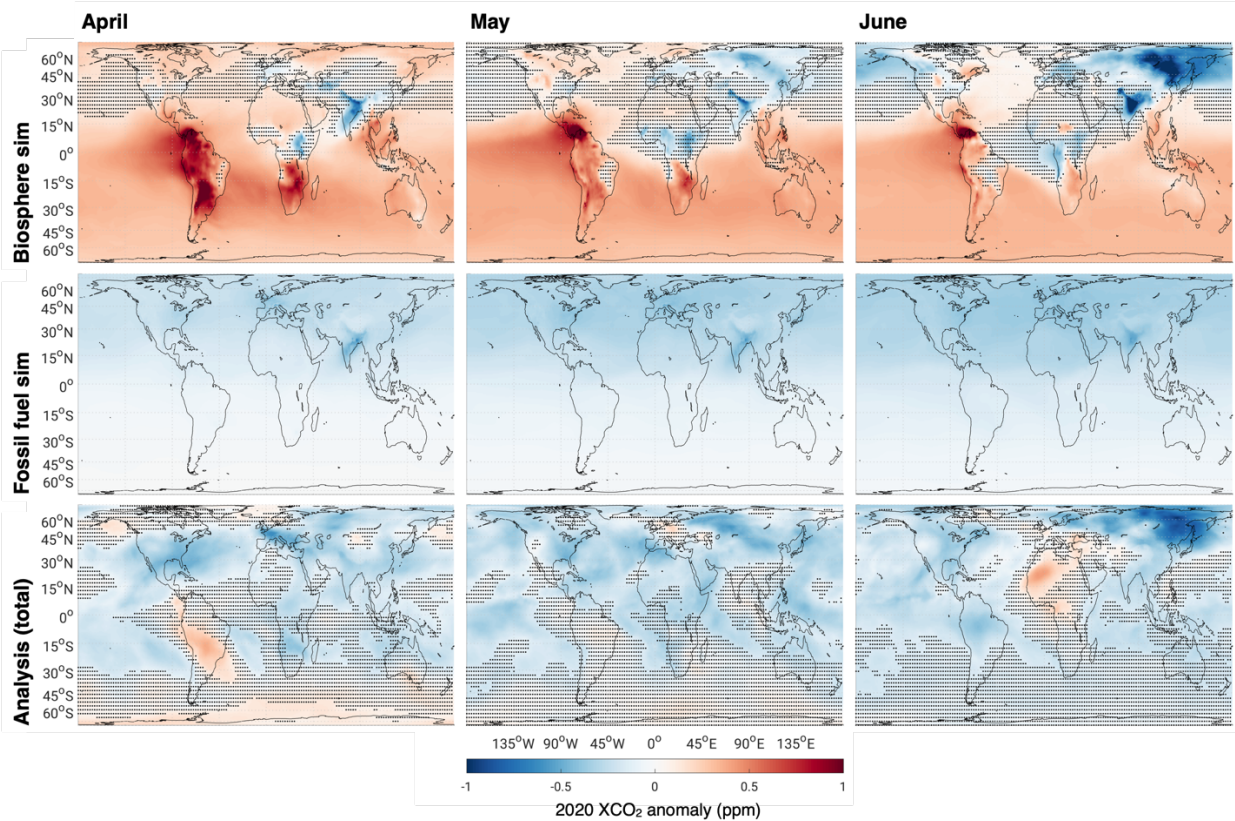

**Fig. S7 (... continued from above and below ...).** In February–May, the simulated biospheric variability over the United States, Europe, Russia, and China simulated remains relatively small and, if anything, net positive, supporting the attribution of the analysis anomalies to fossil fuel emissions. In June, biospheric variability begins to dominate in the Northern Hemisphere, complicating the interpretation of any anthropogenic signals (cf. the growth into June and July of the gray shaded regions in Fig. 3), e.g., the large negative analysis anomaly over Siberia in June (top right) is most likely due to the biosphere (middle right).

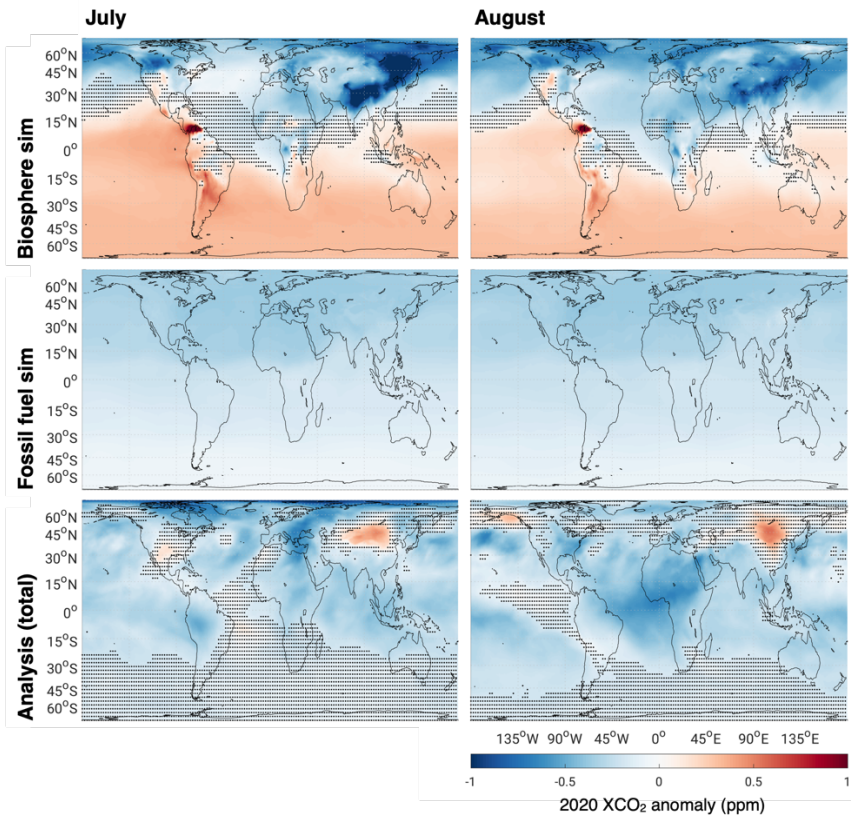

**Fig. S7 (... continued from above).**

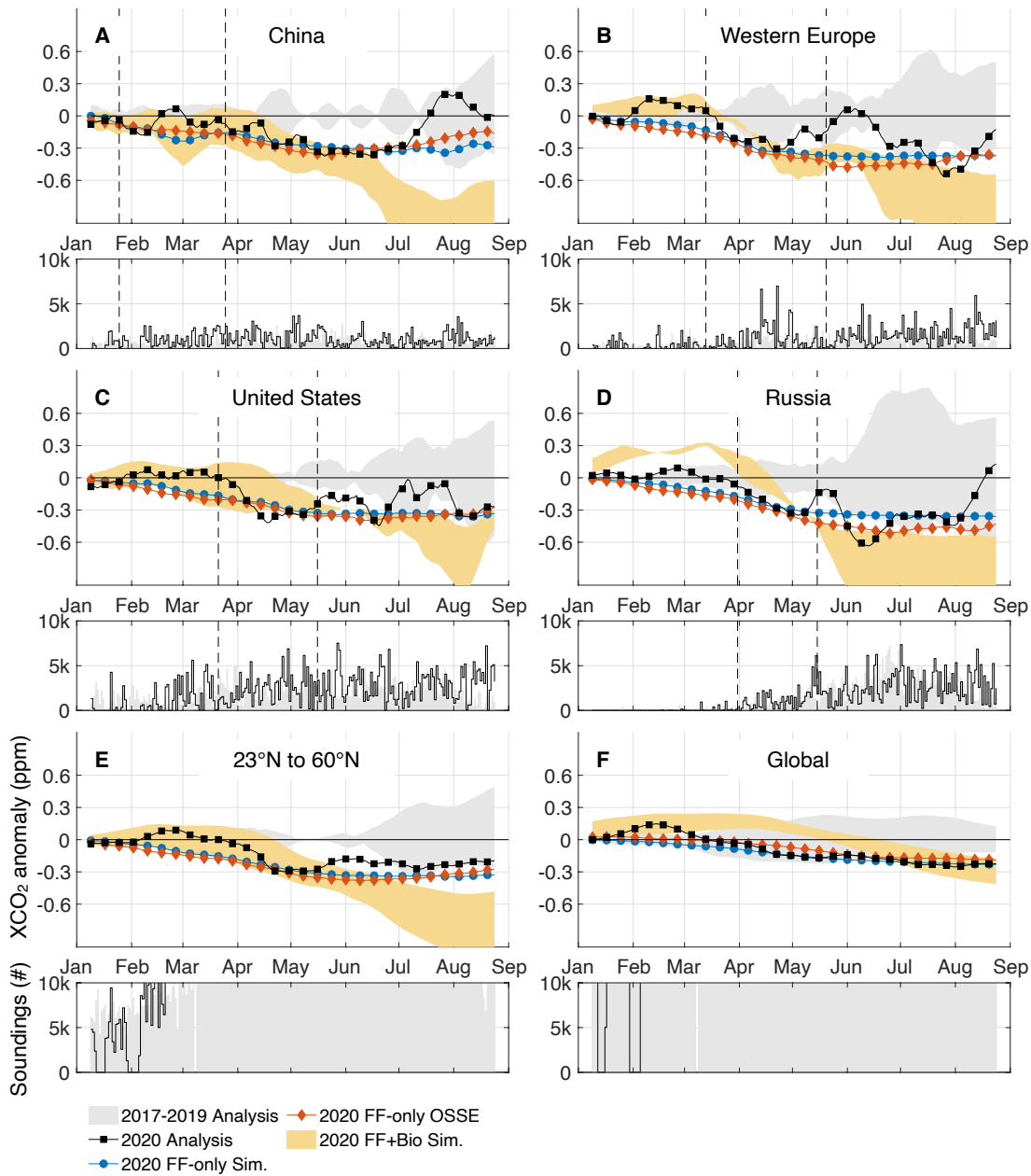

**Fig. S8. Identical to Fig. 3 with fossil fuel OSSE results (red diamonds), fossil fuel plus biosphere anomaly simulations (yellow shading), and daily soundings (lower panel) for 2020 (black line) and the 2017–2019 average (gray shading) included.** The proximity of the fossil fuel OSSE results (red diamonds) to the simulated values (blue circles) demonstrates that the data coverage and assimilation system can reliably capture the signal of activity-based emissions reduction estimates. The distance from the blue circles to the yellow shaded area indicates the magnitude and spread of the simulated biospheric anomalies, and the distance from the black squares to the yellow shaded area the consistency of the analysis anomaly and the simulated anomalies. The agreement of the blue circles, black boxes, and yellow shaded area in the Northern Hemisphere during February–May 2020 supports the conclusion that the assimilation system captured decreases in fossil fuels emissions due to COVID-19.

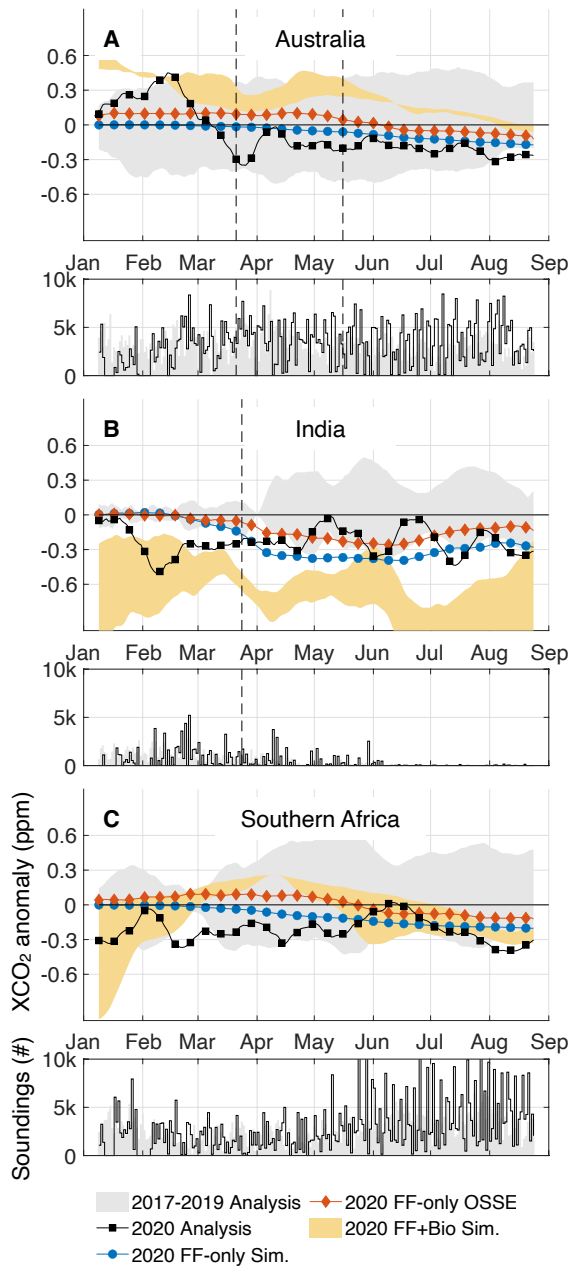

**Fig. S9. Identical to Fig. 5 with OSSE results (red diamonds), fossil fuel plus biosphere anomaly simulations (yellow shading), and daily soundings (lower panel) for 2020 (black boxes) and the 2017–2019 average (gray shading).** While the OSSE results (red diamonds) reproduce the general features of the simulated values (blue circles), the analysis has greater inter-annual variability than in the Northern Hemisphere (compare gray shading in Fig. S8) consistent with a greater biospheric anomaly (distance from blue circles to yellow shaded area). See also the difference between the biospheric (top column) and fossil fuel only (middle column) anomaly maps in the Tropics and Southern Hemisphere in Fig S7.

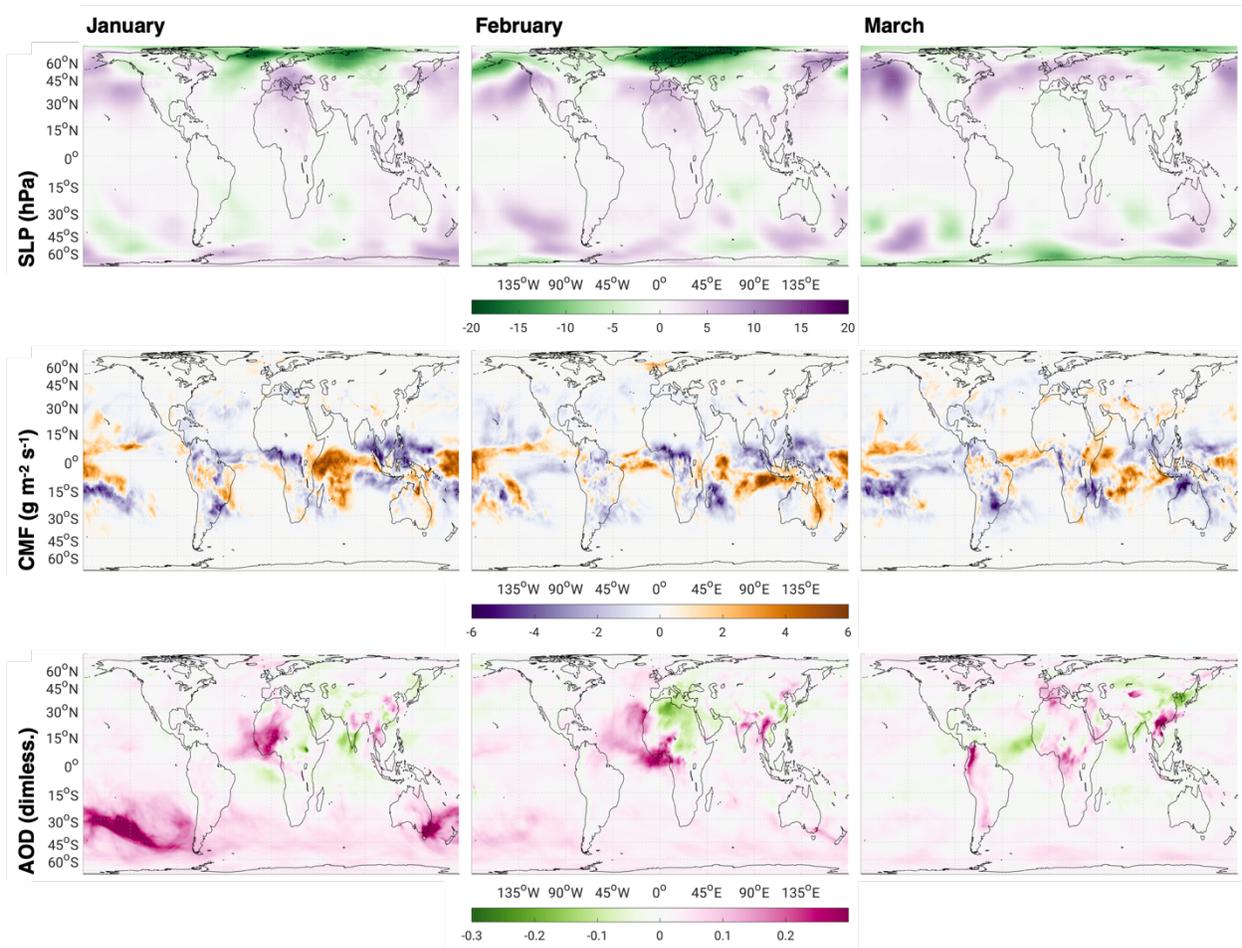

**Fig. S10 (continued below ...). Meteorological anomalies in 2020 for sea-level pressure (SLP; top), convective mass flux (CMF; middle), and aerosol optical depth (AOD) at 755 nm, the Oxygen A-band observed by OCO-2, (bottom) from MERRA-2.** AOD anomalies, which can cause retrieval errors, are relatively small in the Northern Hemisphere during February–May 2020. The remaining plots are included for context.

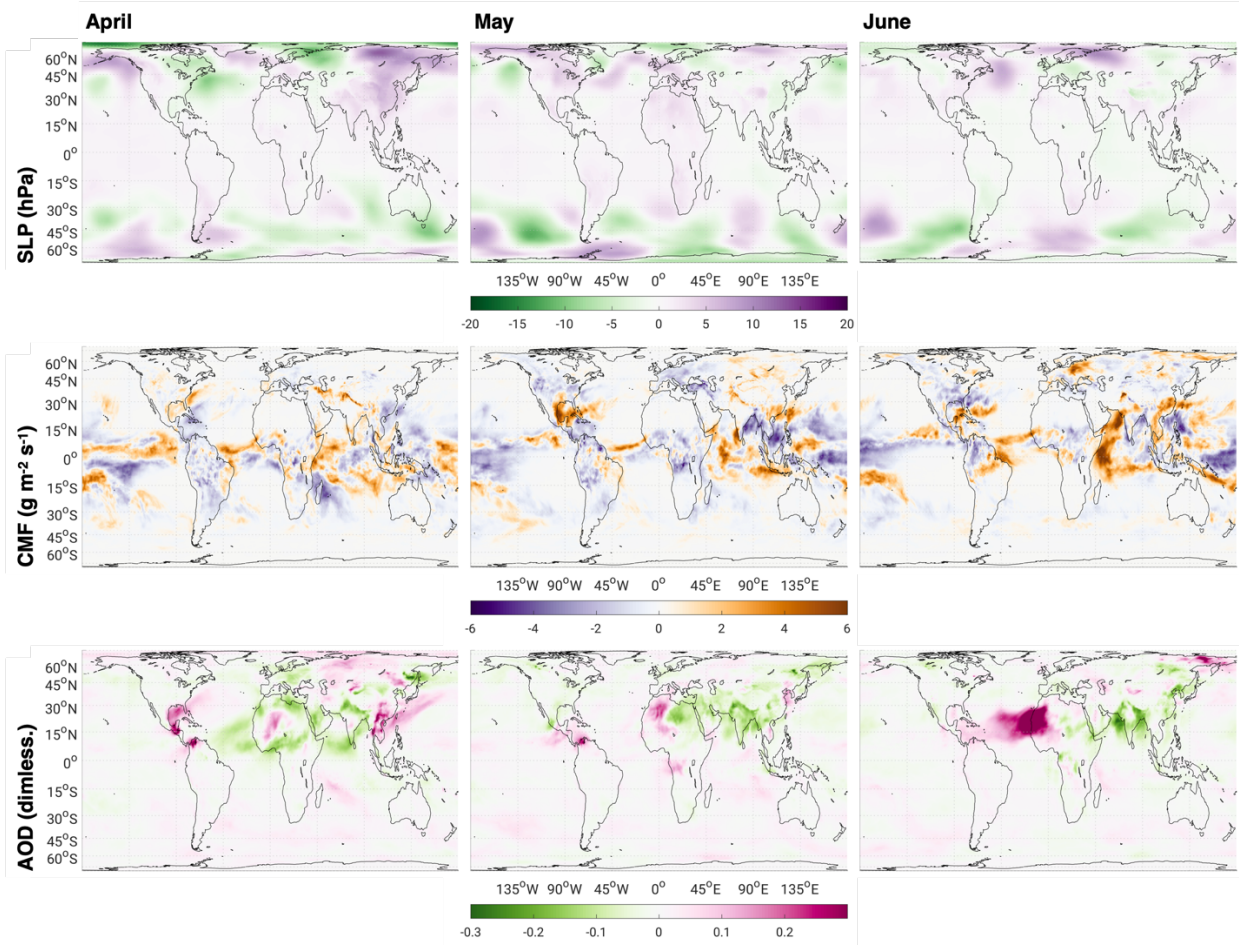

**Fig S10 (... continued from above).**

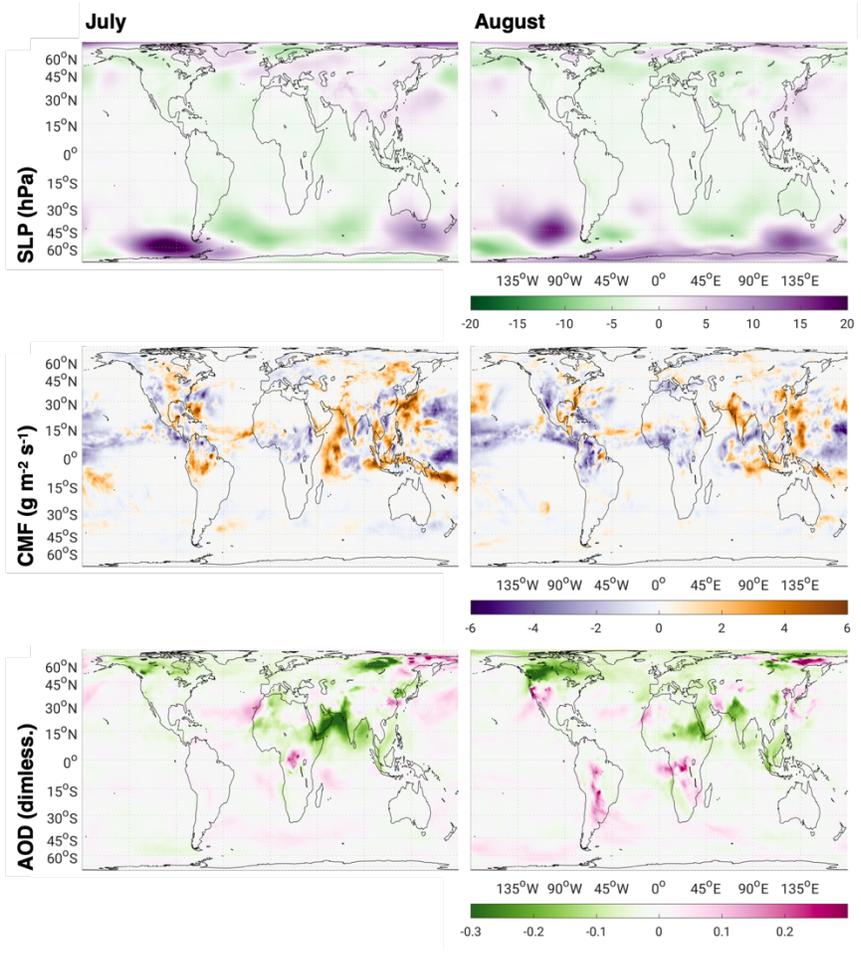

**Fig S10 (... continued from above).**

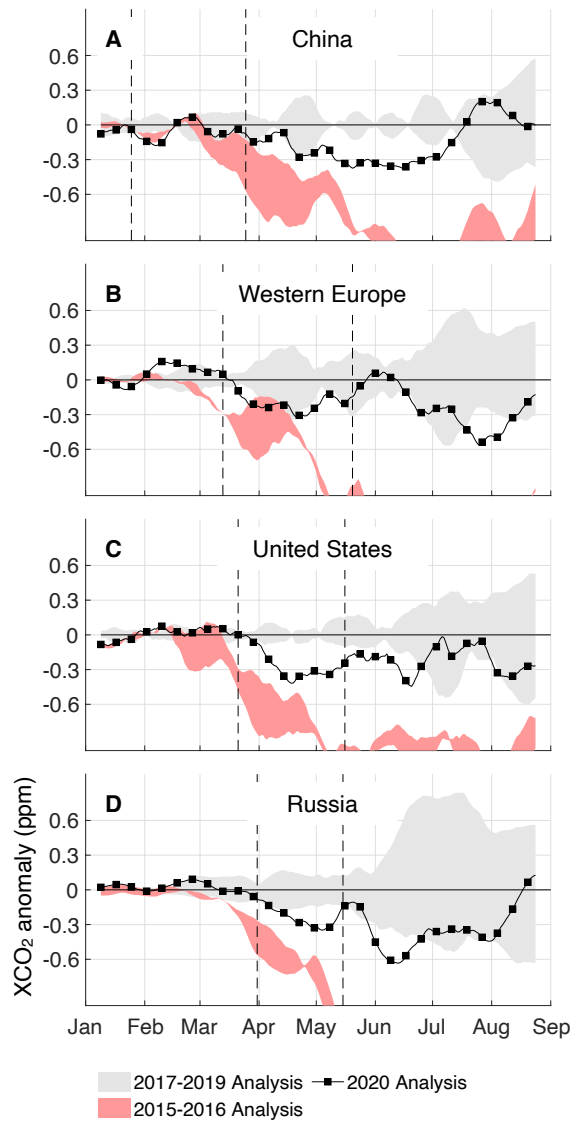

**Fig. S11. Identical to Fig. 3, but with the range of the 2015–2016 ENSO years included.** ENSO anomalies are negative because our surface fluxes underpredict the Northern Hemisphere land sink in these two years. Given the significant differences in the two ranges and the fact that ENSO conditions for 2020 were more reflective of 2017–2019, we exclude 2015–2016 from our analysis.

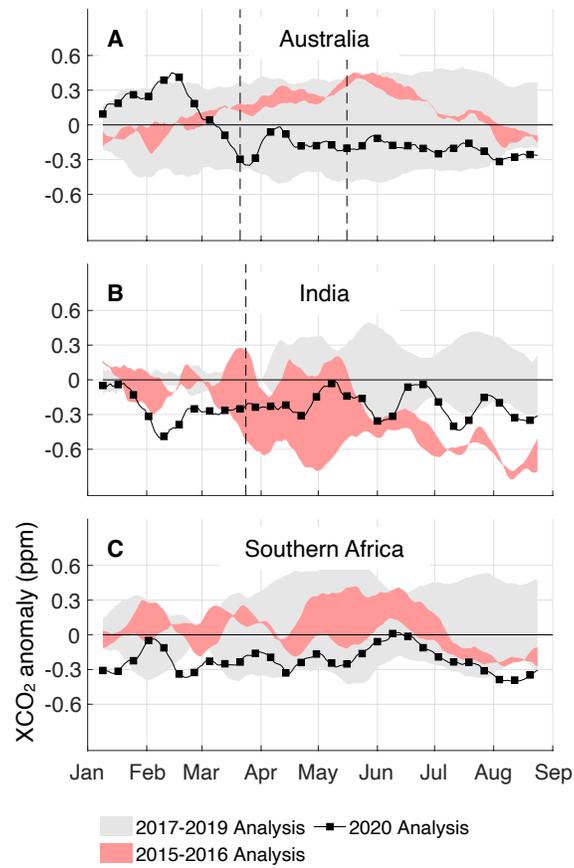

**Fig. S12. Identical to Fig. 5, but with the range of the 2015–2016 ENSO years included.**